\documentclass[%
 preprint,
 amsmath,amssymb,
 aps,
]{revtex4-2}

\usepackage{graphicx}
\usepackage{dcolumn}
\usepackage{bm}
\usepackage{booktabs}
\usepackage{xcolor}
\usepackage{centernot}
\usepackage{appendix}
\usepackage{comment}

\newcommand{\Tp}{T_p}
\newcommand{\ro}{r_0}
\newcommand{\Rep}{Re_p}
\newcommand{\Pep}{Pe_p}
\newcommand{\Rey}{Re}
\newcommand{\St}{St}
\newcommand{\rsup}{\hat{r}_{sup}}
\newcommand{\rrms}{\hat{r}_{rms}}
\newcommand{\boldu}{\mathbf{u}}
\newcommand{\urms}{\hat{U}_{rms}}
\newcommand{\boldup}{\mathbf{u}_p}

\newcommand{\ex}{\hat{\mathbf{e}}_x}

\newcommand{\bu}{\hat{\mathbf{u}}}
\newcommand{\bup}{\hat{\mathbf{u}}_p}

\newcommand{\rhat}{\hat{r}}

\newcommand{\uhat}{\hat{u}}
\newcommand{\vhat}{\hat{v}}
\newcommand{\what}{\hat{w}}

\newcommand{\xihat}{\hat{\xi}}
\newcommand{\Prhat}{\mathcal{P}_{\rhat}}
\newcommand{\Pu}{\mathcal{P}_{\uhat}}
\newcommand{\Pv}{\mathcal{P}_{\vhat}}
\newcommand{\Pw}{\mathcal{P}_{\what}}

\renewcommand{\div}{\nabla \cdot}
\newcommand{\grad}{\nabla}

\begin{document}
\preprint{APS/123-QED}
\title{Preferential concentration by mechanically-driven turbulence in the two-fluid formalism}

\author{Sara Nasab}
 \email{snasab@ucsc.edu}
\author{Pascale Garaud}%
\email{pgaraud@ucsc.edu}
\affiliation{%
Department of Applied Mathematics, Baskin School of Engineering, University of California Santa Cruz, 1156 High Street, Santa Cruz, CA 95064, USA
}%

\begin{abstract}
Preferential concentration is thought to play a key role in promoting particle growth, which is crucial to processes such as warm rain formation in clouds, planet formation, and industrial sprays. In this work, we investigate preferential concentration using 3D Direct Numerical Simulations adopting the Eulerian-Eulerian two-fluid approach, where the particles are treated as a continuum field with its own momentum and mass conservation laws. We consider particles with Stokes number $St \lesssim O(0.01)$ in moderately turbulent flows with fluid Reynolds number $Re \leq 600$. In our previous work (Nasab \& Garaud, \textit{Physical Review Fluids}. doi: 10.1103/PhysRevFluids.5.114308, 2020), we established scaling laws to predict maximum and typical particle concentration enhancements in the context of the particle-driven convective instability. Here we verify that the same results apply when turbulence is externally driven, extending the relevance of our model to a wider class of particle-laden flows. We find in particular that (i) the maximum particle concentration enhancement above the mean scales as $u_{rms}^2 \tau_p / \kappa_p$, where $u_{rms}$ is the rms fluid velocity, $\tau_p$ is the particle stopping time, and $\kappa_p$ is the assumed particle diffusivity from the two-fluid equations; (ii) the typical particle concentration enhancement over the mean scales as $(u_{rms}^2 \tau_p / \kappa_p)^{1/2}$; and (iii) the probability distribution function of the particle concentration enhancement over the mean has an exponential tail whose slope scales like $(u_{rms}^2 \tau_p /\kappa_p)^{-1/2}$. We conclude by discussing the caveats of our model and its  implications in a relevant cloud application.
\end{abstract}
\maketitle
\section{Introduction}\label{sec:intro}
Particle-laden flows are a special class of two-phase fluid flows, characterized by a continuous carrier phase and a dispersed, and typically dilute, particle phase. They appear in numerous physical and engineering applications, including for instance clouds, turbidity currents, protoplanetary disks, and industrial sprays. An important physical process in such flows is the tendency of inertial particles to accumulate in regions of high strain and low vorticity \cite{maxey_1987}, otherwise known as preferential concentration. This process is thought to play a fundamental role in promoting collisional growth. In clouds for example, the growth of micron-size to millimeter size droplets is not quite understood. Although processes such as Brownian motion and condensation can contribute to droplet growth, it is thought that they alone cannot promote sufficient growth to initiate rainfall. For this reason, preferential concentration is considered to be the key process that may result in the enhanced collision rates required for larger raindrop formation \cite{wang2008turbulent, devenish2012droplet, grabowski2013growth}. Similarly in accretion disks, preferential concentration is widely hypothesized to be a vital process for the growth of dust particles into planetesimals \cite{birnstiel2016dust, weidenschilling1993formation}. Thus, our primary goal is to investigate and quantify particle concentration enhancement due to preferential concentration in turbulent flows.

In this work (as also in \citet{nasab2020preferential}), we use the two-fluid formalism and treat the particles as a continuous phase of the system that is distinct from the carrier fluid (see \citet{crowe1996numerical, elghobashi1994predicting, morel2015mathematical} and references within). This continuum approximation is derived
by applying techniques motivated by kinetic theory in which the positions and velocities of the particles are statistically averaged to create a local particle density $\rho_p$ and velocity $\boldup$. We focus on the case where the solid density of the particle $\rho_s$ is much greater than the mean density of the carrier flow $\rho_f$, which is true for many applications. In this limit, the importance of particle inertia is traditionally measured by the Stokes number $\St = \tau_p/\tau_e$, defined as the ratio of the particle stopping time $\tau_p$ to the eddy turnover time $\tau_e$. It has been established that the two-fluid formalism is valid provided that $\St \leq 0.3$ \cite{ferry2002equilibrium}. For larger $\St$, the particles become increasingly uncorrelated with the fluid, and in turn, with one another. When this occurs, the continuum treatment is no longer appropriate.

Using the two-fluid formalism, we recently explored preferential concentration in the context of the particle-induced convective instability \cite{nasab2020preferential}. Our model setup consisted of a carrier fluid with an assumed stable temperature gradient, to which a layer of small and dense inertial particles was added to create linearly unstable initial conditions. We restricted our study to particles with $\St \leq 0.3$ in order to stay within the limit of validity 
of the two-fluid model. In addition, due to the high computational cost required to resolve fine particle structures, we primarily ran 2D Direct Numerical Simulations (DNSs). In all cases, we ran the simulations long enough to study the development of the Rayleigh-Taylor instability, and measured the maximum and typical particle concentration enhancement above the horizontally-averaged particle density. 

Most notably, we found that the maximum particle concentration enhancement above the mean is related to the particle stopping time $\tau_p$, the rms fluid velocity in the turbulent layer $u_{rms}$, and the assumed particle diffusivity $\kappa_p$, scaling as $u_{rms}^2 \tau_p / \kappa_p$. Additionally, we showed that the typical particle concentration enhancement over the mean scales as $(u^2_{\rm{rms}} \tau_p / \kappa_p)^{1/2}$. We also computed the probability distribution function (pdf) of the particle concentration enhancement above the mean and found that in the presence of inertial particles, the tail of the pdf 
appears to be an exponential whose slope scales as $(u^2_{\rm{rms}} \tau_p /\kappa_p)^{-1/2}$. We then explained the importance of the parameter group $u^2_{\rm{rms}} \tau_p / \kappa_p$  using arguments of dominant balance between the inertial concentration and diffusion terms in the particle transport equation (more details can be found in Section \ref{sec:predictive_model}). Although we showed that the model was quite useful in predicting the maximum particle concentration in turbulent flows, our study was limited to flows where the turbulence was driven by the particles themselves. Therefore, whether these results are more generally applicable to any  turbulent particle-laden flow remained to be established. This crucial question is answered in the present paper.

The paper is organized as follows. In Section \ref{sec:model} we introduce the model setup and the governing equations based on the two-fluid formalism. In Section \ref{sec:num_sims} we present DNSs for varying governing parameters (such as the Stokes number and the fluid Reynolds number, for instance), and explore how they affect both preferential concentration and the energetics of the system. In Section \ref{sec:predictive_model} we briefly review the predictive model for maximum particle concentration enhancement presented in \cite{nasab2020preferential} and compare it to the new DNSs. We further look at the typical particle concentration enhancement and the associated pdf of the particle concentration. Section \ref{sec:summary} briefly summarizes and presents applications of our model. We discuss the implications of these results and conclude with final remarks. 

\section{The Model}\label{sec:model}

In this work we use the two-fluid model  described in \citet{nasab2020preferential} to study the dynamics of a dilute suspension of particles in a turbulent carrier fluid. For simplicity, we assume that the  inertial particles have a solid density that is much larger than the mean fluid density such that $\rho_s \gg \rho_f$. We also assume that they are sufficiently small so that Stokes' law can be applied, in which case $\tau_p = \rho_s d_p^2/18\rho_f \nu$, where $d_p$ is the diameter of the particle, $\rho_f$ and $\nu$ are the mean density and the kinematic viscosity of the fluid, respectively. Since the particles are much denser than the fluid, effects incorporated in terms such as the Basset history, Faxen correction, and added mass can be neglected \cite{maxey1983equation}. We additionally require that the particle stopping time $\tau_p$ should be much smaller than the typical eddy turnover time of the carrier fluid $\tau_e$, so $\St \ll 1$.

We use the Boussinesq approximation \cite{boussinesq1903theorie} for the carrier fluid and obtain the following governing equations after a suitable approximation of the particle equations (see \citet{nasab2020preferential}): 
\begin{align}
    &\frac{\partial \boldu}{\partial t} + \boldu \cdot  \grad \boldu  = -\grad p + r \frac{\boldup - \boldu}{\tau_p} + \nu\grad^2\boldu + \frac{1}{\rho_f}\mathbf{F}, \label{eqn:mom} \\ 
    &\frac{\partial \boldup}{\partial t} +  \boldup\cdot\grad\boldup = \frac{\boldu - \boldup}{\tau_p} + \nu_p \grad^2 \boldup, \label{eqn:mom_part} \\
    &\frac{\partial r}{\partial t} + \div (\boldup r) = \kappa_p \grad^2 r,\label{eqn:part} \\ 
    &\div \boldu = 0, \label{eqn:divfree}
\end{align}
where the fluid velocity is $\boldu = (u,v,w)$, $p$ is the pressure, and the particle velocity is $\boldup = (u_p, v_p, w_p)$. Within this formalism, we define the local number density of particles to be $n_p$, and the corresponding mass density to be $\rho_p = n_p m_p$, where $m_p$ is the mass of a single particle. For convenience, we refer to $r = \rho_p/\rho_f$ as the rescaled particle density with respect to the mean density of the carrier fluid (see \citet{nasab2020preferential} for details).

By treating the particles as a continuum, we need to account for the stochastic aspect of particle trajectories, such as Brownian motion and the interaction of a particle with its own or another particle's wake. Generally these interactions are complex in nature, and thus, difficult to implement realistically and numerically. Here for simplicity, we assume that they take the form of a diffusion operator in the equations for the particle density and velocity and set the corresponding diffusivities $\nu_p$ and $\kappa_p$ to be constant. This approximation is valid in the limit where Brownian motion is dominant.

In this work, we drive the turbulence mechanically, by forcing the flow to be shear-unstable. We drive the mean flow using a body force given by $\mathbf{F} = F_0 \sin(k_s z) \ex$, where $F_0$ is the forcing amplitude and $k_s = 2\pi/L_z$ is the wavenumber corresponding to the domain height $L_z$. By selecting a non-cubic domain (where $L_x > L_z$), the Kolmogorov flow thus generated 
is linearly unstable for large enough Reynolds number \cite{beaumont1981stability}. 

We initialize the particles with a uniform distribution in $r$ such that $r=\ro$ everywhere in the domain. In this study, we choose to explore the range $0.1 \leq \ro \leq 10$. Note that $\ro = \Phi_0 \rho_s/\rho_f$, where the initial volume fraction of the particles $\Phi_0$ is small so that the system is well within the dilute limit. Smaller values of $\ro$ correspond to systems such that  $\rho_s/\rho_f \lesssim O(1)$. In this case, particle settling due to gravity is negligible, and can thus be ignored. Larger values of $\ro > 1$ can be obtained when $\rho_s \gg \rho_f$, such as is the case for aerosols or dust in accretion disks. However, particle settling should in principle be taken into account in that limit. Therefore for simplicity, we omit gravity from the particle momentum equation \eqref{eqn:mom_part} to avoid the effect of settling on the dynamics of the system. 


\subsection{Nondimensionalization}\label{subsec:nondim}
In what follows, we define the characteristic length and velocity scales to be
\begin{equation}
    L_c = \frac{1}{k_s} = \frac{L_z}{2\pi} \hspace{.2in} \text{and} \hspace{.2in} U_c = \bigg(\frac{L_zF_0}{2\pi\rho_f}\bigg)^{1/2}, \label{eqn:scales}
\end{equation}
and by construction, the typical eddy turnover time is 
\begin{equation}
    \tau_c = \bigg(\frac{L_z \rho_f}{2\pi F_0} \bigg)^{1/2}. \label{eqn:scales_t}
\end{equation}
This choice effectively assumes a balance in the carrier fluid momentum equation between the inertial terms and the forcing, such that $\boldu \cdot \grad \boldu \sim F_0 /\rho_f$. After using \eqref{eqn:scales} and \eqref{eqn:scales_t} to scale Eqs. \eqref{eqn:mom}-\eqref{eqn:divfree}, the nondimensional governing equations are
\begin{align}
    &\frac{\partial \bu}{\partial t} + \bu \cdot  \grad \bu  = -\grad \hat{p} + r_0\rhat \frac{\bup - \bu}{\Tp} + \frac{1}{\Rey}\grad^2\bu + \sin(z)\ex , \label{eqn:nd_mom} \\
    &\frac{\partial \bup}{\partial t} +  \bup\cdot\grad\bup = \frac{\bu - \bup}{\Tp} + \frac{1}{\Rep} \grad^2 \bup, \label{eqn:nd_mompart} \\
    &\frac{\partial \rhat}{\partial t} + \div (\bup \rhat) = \frac{1}{\Pep} \grad^2 \rhat,\label{eqn:nd_part} \\ 
    &\div \bu = 0, \label{eqn:nd_divfree}
\end{align}
where the hatted quantities (as well as the independent variables) are now nondimensional, where $\rhat=r/r_0$, and where   
\begin{equation}
    \Tp = \frac{\tau_p}{\tau_c}
\end{equation} 
is the nondimensional stopping time, which can be viewed as a first estimate of the Stokes number. Additionally, the diffusion terms are now characterized by a Reynolds number for the fluid $\Rey$, a Reynolds number for the particles $\Rep$, and the particle P\'eclet number $\Pep$ respectively defined by
\begin{equation}
    \Rey = \frac{U_c L_c}{\nu}, \hspace{.8cm} \Rep = \frac{U_c L_c}{\nu_p}, \hspace{.8cm} \Pep = \frac{U_c L_c}{\kappa_p}.
\end{equation}

\section{Numerical simulations}\label{sec:num_sims}

\subsection{The PADDI-2F code}

We use Direct Numerical Simulations to investigate the effects of preferential concentration in the model described in Section \ref{sec:model}. We use a modified version of the pseudospectral PADDI code, which was originally developed to study double-diffusive phenomena in oceanic contexts \cite{traxler2010dynamics, traxler2011numerically, stellmach2010dynamics}, and later extended to astrophysical applications \cite{moll2016new, garaud2016turbulent} and to particle-laden flows \cite{nasab2020preferential}. PADDI-2F solves the governing equations \eqref{eqn:nd_mom}-\eqref{eqn:nd_divfree} in spectral space. Specifically, diffusion terms are treated implicitly in spectral space, whereas both nonlinear and drag terms are first computed in real space, transformed into spectral space, and then, integrated explicitly using a third-order Adams-Bashforth backward-differencing scheme. Drag terms are computed in a way that ensures the total momentum is conserved (other than the dissipation terms) throughout the simulations.

The computational domain is triply-periodic, with $(L_x, L_y, L_z) = (4\pi, 2\pi, 2\pi)$ to ensure that the flow is linearly unstable under the selected forcing. All simulations are run until a statistically steady state has been reached, either starting from the initial conditions as described in Section \ref{sec:model}, or starting from the end of another simulation at nearby parameters. Due to the high cost of running simulations in 3D and the resolution needed to resolve fine-scale particle structures, we restrict our simulations to  $\Rey \leq 600$ and up to moderate values of $\Tp \leq 0.03$ in which the two-fluid formalism is valid. Specifications of all simulations are listed in Table \ref{table:sims}.

\subsection{The effect of $\Rey$ on turbulence}\label{subsec:compare_Rey_nopart}

We first look at the influence of $\Rey$ on the turbulence 
in the absence of particles, which will be used as a reference point for later simulations with particles. We therefore only use the momentum equation \eqref{eqn:nd_mom} and the divergence-free condition \eqref{eqn:nd_divfree}, and set $\ro = 0$. We set the resolution of the 3D runs to be $768 \times 384 \times  384$ equivalent grid points in the $x-, y-$, and $z-$directions, respectively.

\begin{figure}
        \centering
        \includegraphics[width=0.5\textwidth]{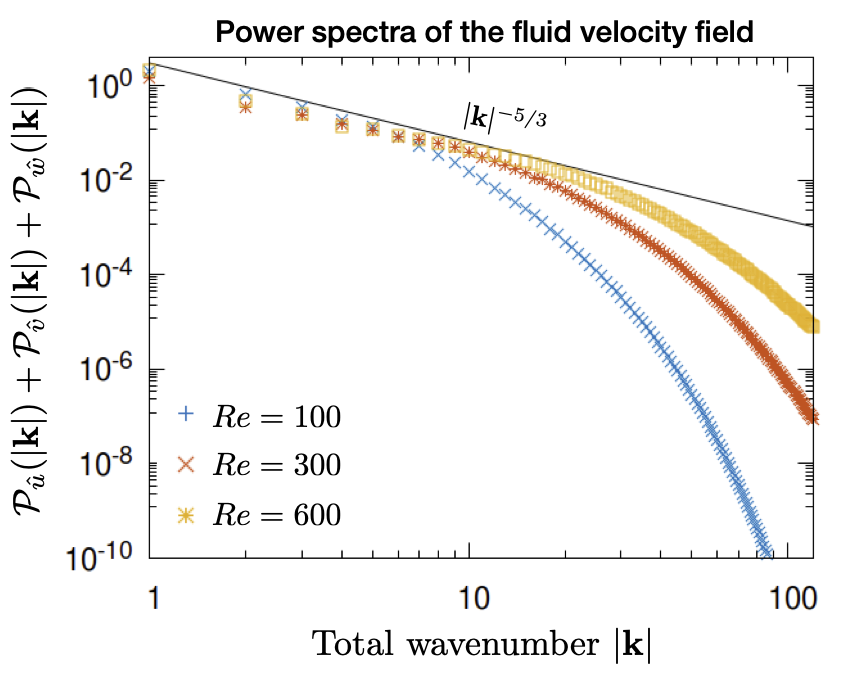} 
        \caption{Instantaneous power spectra 
        of the total fluid velocity field as function of $|\mathbf{k}|$ for simulations in the absence of particles for $\Rey = 100$, $300$, and $600$. The black solid line scales as $|\mathbf{k}|^{-5/3}$.  }
        \label{fig:spectra_Rey}
\end{figure}

We examine the power spectra of the fluid velocity field once the system has reached a statistically steady state, 
and compare the results for different Reynolds numbers. We define the 
power in mode $\mathbf{k}$ 
for a scalar quantity $\hat{\xi}$ (e.g. $\hat{u}, \hat{v},$ and $\hat{w}$) as
\begin{equation}\label{eqn:power_spec}
    P_{\hat{\xi}}(\mathbf{k}, t) = \tilde{\xi}(\mathbf{k},t) \tilde{\xi}^*(\mathbf{k},t)
\end{equation}
where $\mathbf{k}=(k_x,k_y,k_z)$ is the wavevector and $\tilde{\xi}(\mathbf{k},t)$ and $\tilde{\xi}^*(\mathbf{k},t)$ are the Fourier transform of $\hat{\xi}$ and its complex conjugate, respectively.  For Figures \ref{fig:spectra_Rey}, \ref{fig:spectra_varyTp}, \ref{fig:spectra_varyr0}, \ref{fig:spectra_varyRef_Tp0.01}, and  \ref{fig:spectra_Pep} we present the power spectra $\mathcal{P}_{\xihat}(|\mathbf{k}|, t)$ as a function of the total wavenumber $|\mathbf{k}| = (k_x^2 + k_y^2 + k_z^2)^{1/2}$, where $\mathcal{P}_{\xihat}(|\mathbf{k}|, t)$ is the power contained in all the modes whose amplitudes lie between 
$|\mathbf{k}|$ and $|\mathbf{k}| + 1$. 

Figure \ref{fig:spectra_Rey} presents the power spectra of the total fluid velocity field $\mathcal{P}_{\uhat}(|\mathbf{k}|) + \mathcal{P}_{\vhat}(|\mathbf{k}|) + \mathcal{P}_{\what}(|\mathbf{k}|)$ extracted at an instant in time after the system has reached a statistically steady state for three simulations with $\Rey = 100, 300$, and 600, respectively. For sufficiently large $\Rey$, the system exhibits a well-known energy cascade whose inertial range scales as $|\mathbf{k}|^{-5/3}$, shown
here by the black line for ease of comparison. As expected, we find that the inertial range increases with $\Rey$ and ends at the Taylor microscale $\lambda = \sqrt{15}\Rey^{-1/2}L_z$. This corresponds to $k_{\lambda} = 2\pi/\lambda$, which is equal to $k_{\lambda} \approx 2.6$ for $\Rey = 100$, $k_{\lambda} \approx 4.5$ for $\Rey  = 300$, and $k_{\lambda} \approx 6.3$ for $\Rey  = 600$. In geophysical and astrophysical applications, $\Rey$ is much larger, with an established inertial range spanning many orders of magnitude, which we do not see for the simulations presented here. Therefore, one must be careful about extrapolating the results obtained in this paper to systems with $\Rey \gg 10^3$ (see Section \ref{subsec:discussion} for more details).

\subsection{The effect of $\Tp$ on preferential concentration}\label{subsec:compare_tp}

We now explore how the other input parameters
affect preferential concentration, and how this in turn alters the energetics of the system. We first look at how the non-dimensional stopping time, which is also a proxy for the Stokes number of the particles, affects the system by comparing a 3D simulation with a very low $\Tp = 0.005$ to another at a higher $\Tp = 0.03$. To do so, we use the PADDI-2F code using Eqs. \eqref{eqn:nd_mom}--\eqref{eqn:nd_divfree}, with the remaining parameters set as $\ro = 0.1, \Rey = 100, \Rep = 600$, and $\Pep = 600$. The resolution and domain size for the simulations are set to $768 \times 384 \times  384$ equivalent grid points, and as before, $L_x = 4\pi$ and $L_y=L_z = 2\pi$ (see Table \ref{table:sims} for more details). 

\begin{figure}
        \includegraphics[width=1\textwidth]{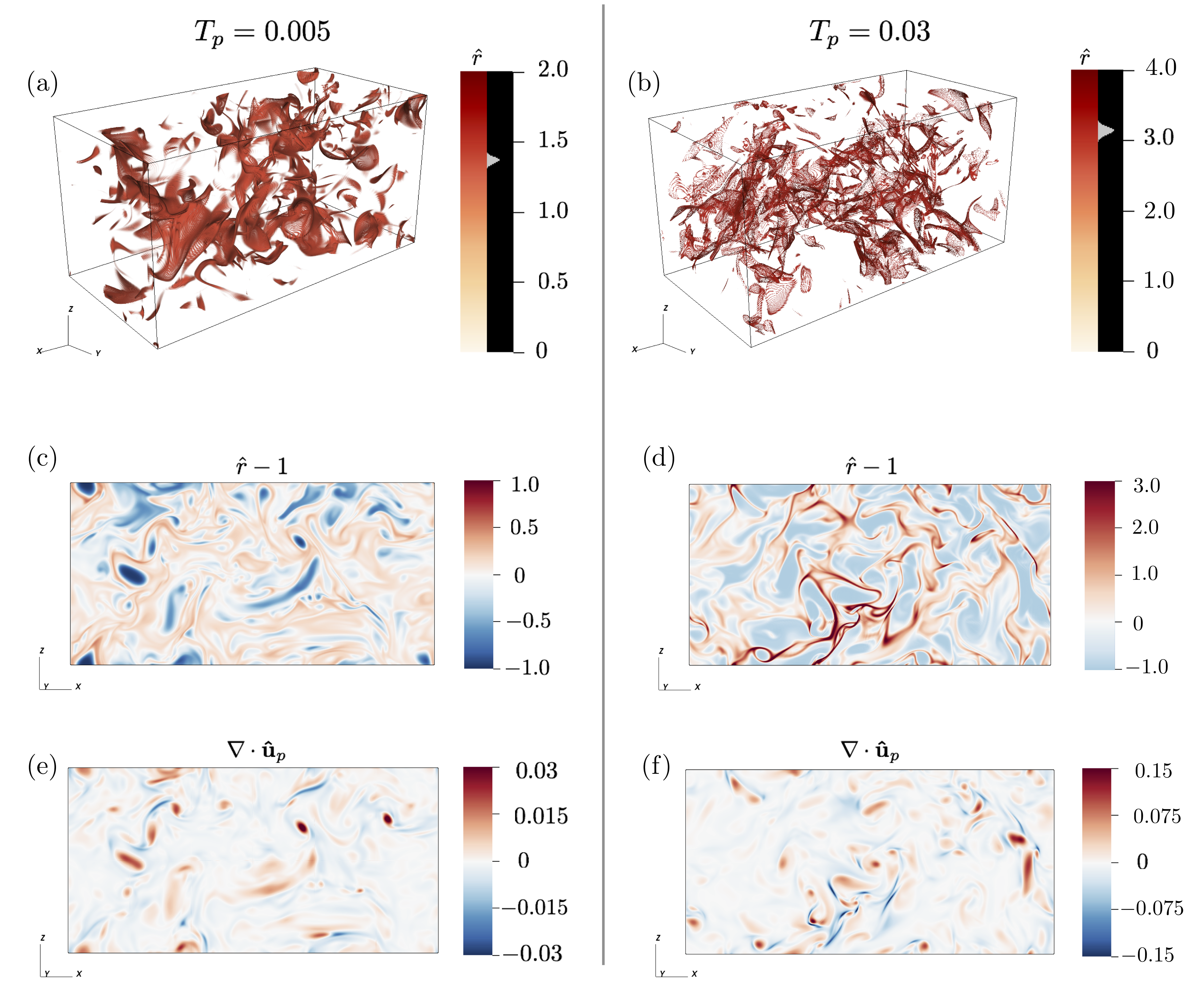}
        \caption{Comparison of particle concentration snapshots for low $\Tp = 0.005$ (left column) and high $\Tp = 0.03$ (right column) simulations. Each snapshot was extracted once the system has reached a statistically steady state. (a)-(b): volume rendering of $\rhat$; (c)-(d): snapshots of the particle concentration enhancement $\rhat- 1$ at $y = 0$; (e)-(f): snapshots of $\div \bup$ at $y = 0$. The remaining parameters are: $\ro = 0.1, \Rey = 100, \Rep = 600, \Pep = 600$.}
        \label{fig:snaps}
\end{figure}

We present snapshots in  Figure \ref{fig:snaps} of the particle concentration field after the system has reached a statistically steady state. In the volume renderings shown in   
Figures \ref{fig:snaps}(a) and \ref{fig:snaps}(b), we highlight areas of relatively higher particle concentration in red. Both simulations consist of sheet-like particle structures which appear to be about the same size, but denser for the high $\Tp$ case. We can see the particle structures in more detail in  Figures \ref{fig:snaps}(c) and \ref{fig:snaps}(d), which show
the particle concentration deviation from the mean (namely, $\rhat -1$) in 
a slice taken at $y=0$. We clearly see that the denser particle structures indeed appear to be the same size for both simulations. The densest structures for the high $\Tp$ case have values of 
$\rhat - 1 \approx 3$ compared to $\rhat - 1 \approx 0.5$ for structures found in the low $\Tp$ case. 

The fact that preferential concentration is more efficient at higher values of $\Tp$ recovers the well-known results of \cite{maxey_1987}, which are expressed as follows in the two-fluid formalism. Using the particle momentum equation \eqref{eqn:nd_part}, we can express $\bup$ in terms of $\bu$ and $\Tp$ using an asymptotic expansion in $\Tp$: 
\begin{equation}\label{eqn:up_asymp}
	\bup = \bu - \Tp \bigg(\bu\cdot \grad \bu + \frac{\partial \bu}{\partial t} - \frac{1}{\Rep}\grad^2 \bu \bigg) + O(T_p^2). 
\end{equation}
Taking the divergence of \eqref{eqn:up_asymp}, we then obtain 
\begin{equation}\label{eqn:divup}
	\div \bup = - \Tp \div (\bu \cdot \grad \bu) + O(\Tp^2),
\end{equation}
which shows that 
$\div \bup$ is non zero even though $\div \bu= 0$, and furthermore depends linearly on $\Tp$ for small $\Tp$. It is easy to see (from Eq. \ref{eqn:nd_part}) that the particle concentration grows (or decays)  exponentially since
\begin{equation}
    \frac{\partial \rhat}{\partial t} = - \rhat (\div \bup) + ...,
\end{equation}
showing that the growth or decay rate of $\rhat$
is given by $|\div \bup|$. Figure \ref{fig:snaps} compares the particle concentration enhancement $\rhat-1$ (panels \ref{fig:snaps}(c) and \ref{fig:snaps}(d)) to the value of $\div \bup$ (panels \ref{fig:snaps}(e) and \ref{fig:snaps}(f)) at the same time. We see that areas where $\div \bup < 0$ (shown in blue) correspond to regions where $\rhat -1$ is maximal, while regions with $\div \bup > 0$ (shown in red) correspond to regions where $\rhat$ is close to 0 (equivalently, $\rhat-1$ is close to $-1$), as expected from the argument above.


\begin{figure}
        \includegraphics[width=1\textwidth]{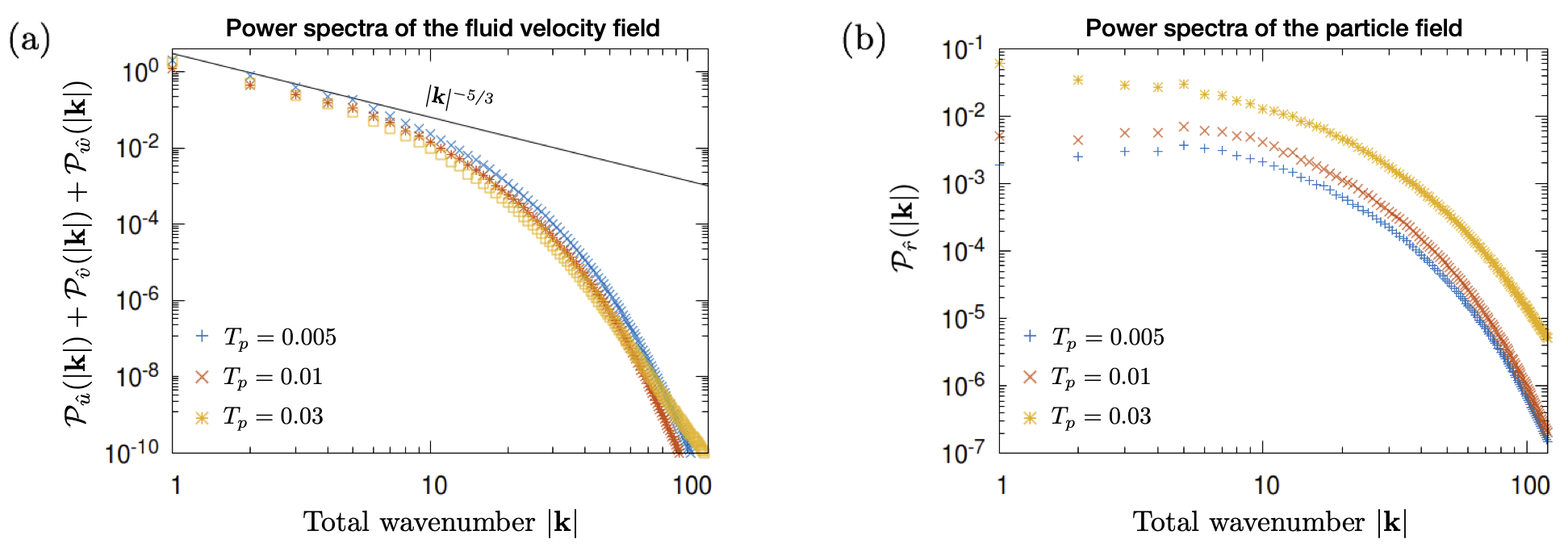} 
        \caption{Instantaneous power spectra of (a) the total fluid velocity field and (b) the particle concentration field as a function of the total wavenumber $|\mathbf{k}|$ for varying $\Tp$. The remaining parameters are $\ro = 0.1, \Rey = 100, \Rep = 600,$ and $\Pep = 600$.   }
        \label{fig:spectra_varyTp}
\end{figure}

We next compare the power spectra (using Eq. \ref{eqn:power_spec})  of the particle concentration and fluid velocity fields for simulations with varying $\Tp$, with the remaining parameters fixed as $\ro = 0.1, \Rey = 100, \Rep = 600$ and $\Pep=600$. Figure \ref{fig:spectra_varyTp}(a) shows the power spectrum of the total fluid velocity field $\mathcal{P}_{\uhat}(|\mathbf{k}|) + \mathcal{P}_{\vhat}(|\mathbf{k}|) + \mathcal{P}_{\what}(|\mathbf{k}|) $. 
The solid black line represents the Kolmogorov spectrum given by $|\mathbf{k}|^{-5/3}$. Although there is a subtle decrease in power across all scales for larger $\Tp$, the velocity spectrum appears to be overall unaffected. 

Figure \ref{fig:spectra_varyTp}(b) shows the power spectrum of the particle concentration $\Prhat(|\mathbf{k}|)$. 
We see that increasing $\Tp$ causes an increase in $\Prhat(|\mathbf{k}|)$ at all scales, with the exception of the $\mathbf{k} = \mathbf{0}$ mode whose amplitude  
instead decreases (not shown here). This is consistent with our expectation that increasing inertia causes an increase in preferential concentration, and is directly related to the
snapshots in Figure \ref{fig:snaps}: comparable-sized particle structures are denser (higher $\rhat$) for large $\Tp$ than for small $\Tp$. 

\subsection{The effect of $\ro$ on preferential concentration}\label{subsec:compare_ro}
\begin{figure}
        \includegraphics[width=1\textwidth]{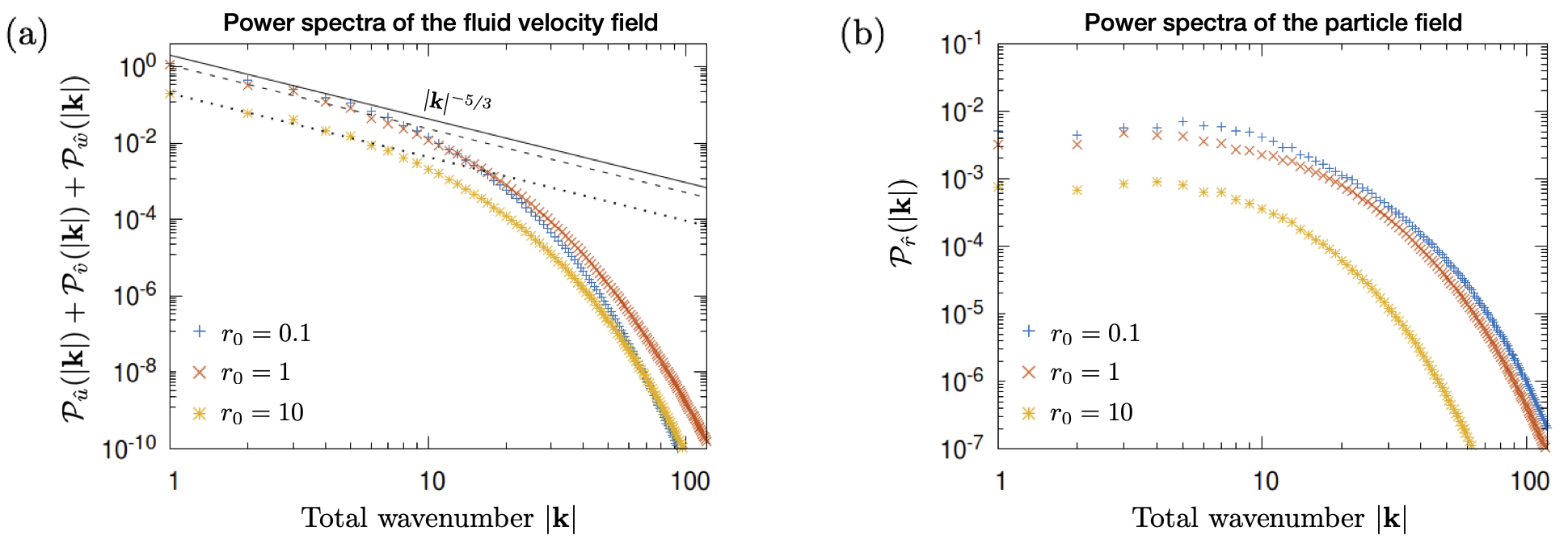} 
        \caption{Instantaneous power spectra of (a) the total fluid velocity field and (b) the particle concentration field as a function of the total wavenumber $|\mathbf{k}|$ for varying $\ro$. The remaining parameters are $\Tp = 0.01, \Rey = 100, \Rep = 600$, and $\Pep = 600$.  The solid, dashed, and dotted lines represents the predicted scaling for the power at the injection scale for $\ro = 0.1, 1,$ and $10$, respectively (see main text for details).}
        \label{fig:spectra_varyr0}
\end{figure}

We next look at the effect of the particle mass loading fraction $\ro$ on the energetics of the system. We set $\Tp = 0.01, \Rey = 100, \Rep = 600,$ and $\Pep=600$ (see Table \ref{table:sims} for more details) and present the power spectra of the particle concentration and total fluid velocity fields (using Eq. \ref{eqn:power_spec}) in  Figures \ref{fig:spectra_varyr0}(a) and \ref{fig:spectra_varyr0}(b), respectively. 

Turning first to the power spectrum of the velocity field in  Figure \ref{fig:spectra_varyr0}(a), we see that $\ro$ has a strong effect on the total energy of the turbulent flow (or equivalently, on the normalization of the power spectrum). 
The inertial range still shows the usual $|\mathbf{k}|^{-5/3}$ power law, and ends around $k_{\lambda} \approx 2.6$ in all three simulations, which is the same as the case without particles ($\ro = 0$ with $\Rey = 100, \Rep = 600, \Pep=600$ in  Figure \ref{fig:spectra_Rey}). In order to understand why increasing $\ro$ reduces the turbulent energy, note that in a statistically stationary state, the momentum equation \eqref{eqn:mom} reaches a balance between the inertial terms and body force terms, expressed dimensionally as $\rho_f (\boldu \cdot \grad \boldu) \sim \mathbf{F}$. With the addition of particles that are well-coupled to the fluid, the dominant balance becomes $(\rho_f + \rho_p) (\boldu \cdot \grad \boldu) \sim \mathbf{F}$. In the nondimensionalization presented in Section \ref{subsec:nondim}, this balance implies $(1+\ro){\bu}^2 \simeq O(1)$. With this in mind, we can then expect that $\Pu(|\mathbf{k}|) + \Pv(|\mathbf{k}|) + \Pw(|\mathbf{k}|)$ ought to scale like $1/(1+\ro)$ at the injection scale. The scaling is confirmed in  Figure \ref{fig:spectra_varyr0} for $\ro = 0.1, \ro =1, $ and $\ro = 10$ given by the solid, dashed, and dotted lines, respectively.

Figure \ref{fig:spectra_varyr0}(b) shows the power spectrum of the particle concentration field $\Prhat(|\mathbf{k}|)$, 
and we see that larger $\ro$ corresponds to smaller $\Prhat(|\mathbf{k}|)$ across all scales (except the $\mathbf{k} = \mathbf{0}$ mode which is not shown). This is consistent with the fact that larger $\ro$ results in a decrease in the turbulence intensity (and therefore preferential concentration) across all scales, as observed from the velocity power spectrum.

\subsection{The effect of $\Rey$ on preferential concentration}\label{subsec:compare_Rey}

\begin{figure}
        \includegraphics[width=1\textwidth]{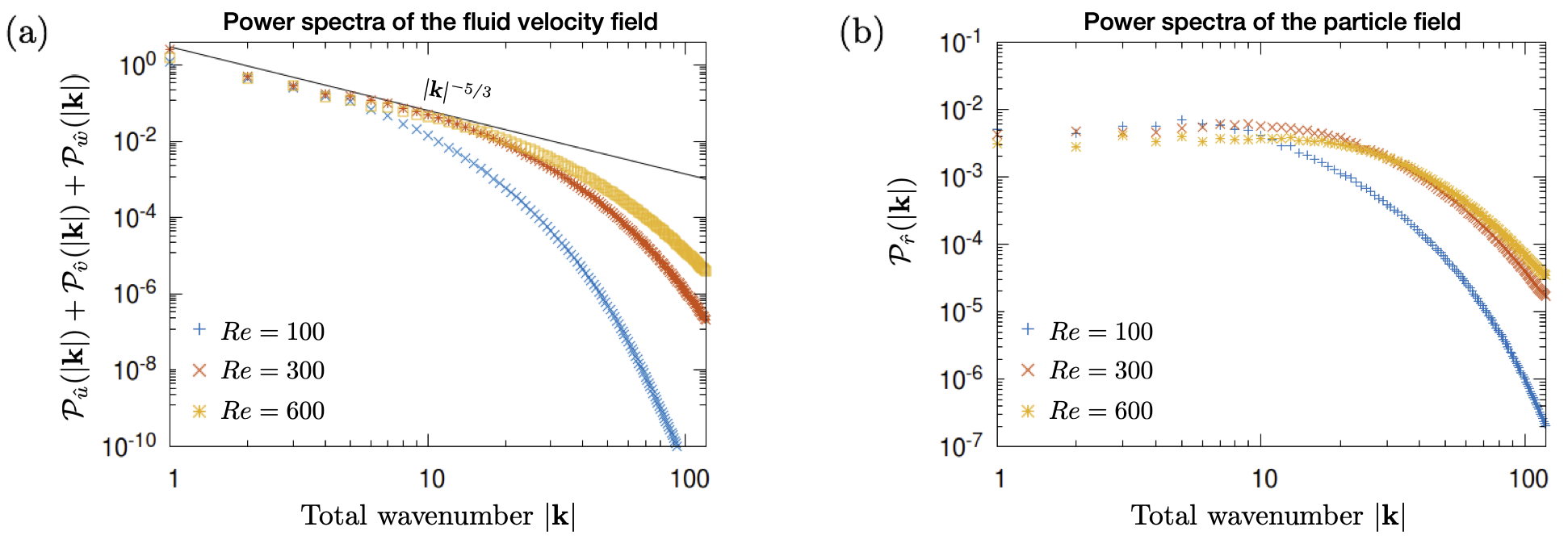} 
        \caption{Instantaneous power spectra of (a) the total fluid velocity field and (b) the particle concentration field as a function of the total wavenumber $|\mathbf{k}|$ for varying $\Rey$. The remaining parameters are $\Tp = 0.01, \ro = 0.1, \Rep = 600$, and $\Pep = 600$. }
        \label{fig:spectra_varyRef_Tp0.01}
\end{figure}

In this section, we investigate how varying the fluid Reynolds number affects the energetics, while fixing the other parameters to be $\Tp = 0.01$, $\ro = 0.1, \Rep = 600$, and $\Pep = 600$ (see Table \ref{table:sims} for more details). Figures \ref{fig:spectra_varyRef_Tp0.01}(a) and \ref{fig:spectra_varyRef_Tp0.01}(b) show the power spectra of the total fluid velocity and particle concentration fields, respectively (using Eq. \eqref{eqn:power_spec}). 

\begin{figure}
        \centering
        \includegraphics[width=\textwidth]{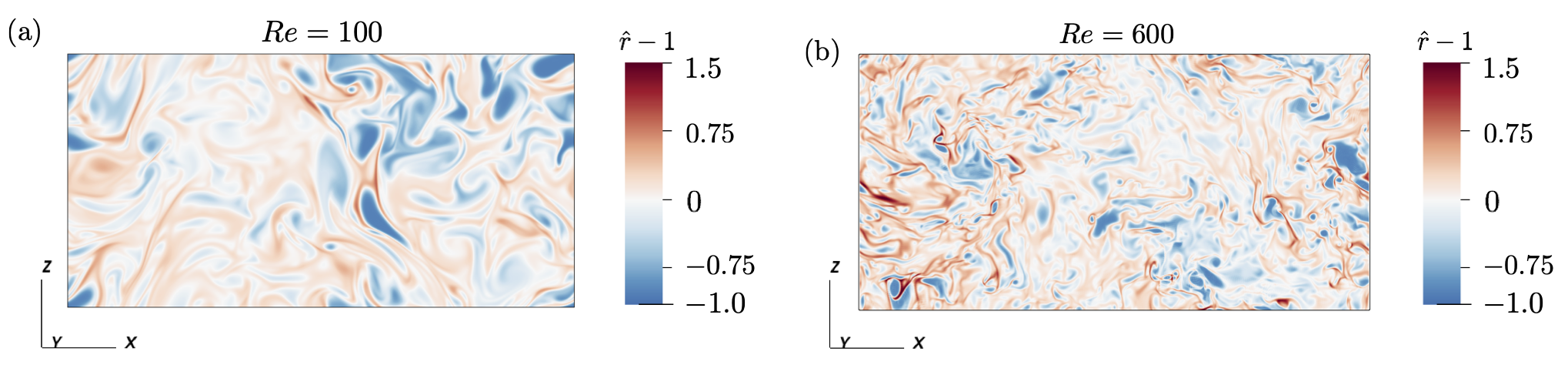} 
        \caption{Snapshots of the particle concentration enhancement above the mean $\rhat - 1$ for two different simulations with $\Rey = 100$ and $\Rey = 600$. The remaining parameters are $\Tp = 0.01, \ro = 0.1, \Rep = 600$, and $\Pep = 600$.}
        \label{fig:snaps_Rey}
\end{figure}

In  Figure \ref{fig:spectra_varyRef_Tp0.01}(a), the 
velocity spectra shown are more or less indistinguishable from those of the corresponding fluid-only simulations presented in Section \ref{subsec:compare_Rey_nopart}. This is not surprising since the value of $\ro = 0.1$ chosen for these simulations is quite small. Because a larger $\Rey$ extends the Kolmogorov cascade, finer scales of turbulence are generated. Consequently in  Figure \ref{fig:spectra_varyRef_Tp0.01}(b), we also see substantially more power in the particle density field at smaller scales when $\Rey$ increases.  This is also confirmed in the snapshots of the particle concentration enhancement ($\rhat - 1$) shown in  Figure \ref{fig:snaps_Rey}, which compare simulations with 
$\Rey = 100$ and $\Rey = 600$. The 
finer scales of turbulence for larger $\Rey$ cause the denser particle structures to appear overall more fragmented and convoluted. 
On the other hand, we note that the densest filamentary regions (shown in dark red) have comparable 
thickness for varying $\Rey$. 

\subsection{The effect of $\Pep$ and $\Rep$ on preferential concentration}\label{subsec:compare_Pep_Rep}

Finally, we examine how the particle diffusion coefficients 
$\Pep$ and $\Rep$ affect the energetics of the system. As described in Section \ref{sec:model}, the particle concentration and momentum diffusivities are necessary when modeling the particles as a continuum, but their origin is grounded in the notion that the particle velocities have a stochastic component in addition to the mean $\bup$. Since the origin of the  particle diffusivity is likely the same as that of the momentum diffusivity, we may expect $\Pep$ and $\Rep$ to be related, and close to one another. In what follows, we take $\Pep = \Rep$ for simplicity. 

\begin{figure}
        \includegraphics[width=1\textwidth]{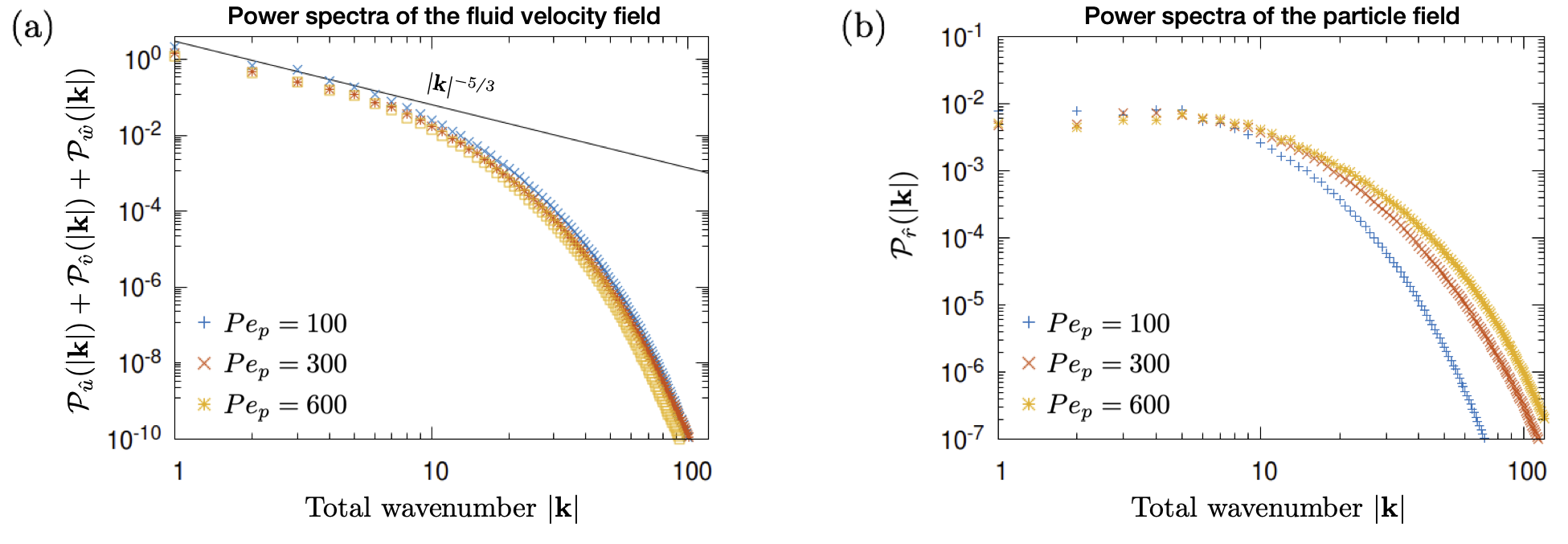} 
        \caption{Instantaneous power spectra of (a) the total fluid velocity field and (b) the particle concentration field as a function of the total wavenumber $|\mathbf{k}|$ for varying $\Pep$. 
        The remaining parameters are $\Tp = 0.01, \ro = 0.1$, $\Rey = 100$, and $\Rep=\Pep$. }
        \label{fig:spectra_Pep}
        \end{figure}

Figure \ref{fig:spectra_Pep} shows the power spectra of the particle concentration field and the total fluid velocity field. In  Figure \ref{fig:spectra_Pep}(b), we see that a larger $\Pep$  (equivalently, a lower particle diffusivity) results in significantly larger $\Prhat(|\mathbf{k}|)$ across all scales, and thus, the presence of smaller-scale structures in the particle concentration field. By contrast, we see in  Figure \ref{fig:spectra_Pep}(a) that $\Pep$ does not affect the velocity power spectrum significantly, other than a slight decrease in energy across all scales for larger $\Pep$.

\begin{table}
  \begin{center}
  \begin{tabular}{c@{\hspace{.4cm}}c@{\hspace{.4cm}}c@{\hspace{.4cm}}c@{\hspace{.4cm}}c@{\hspace{.4cm}}c@{\hspace{.4cm}}c@{\hspace{.4cm}}c@{\hspace{.4cm}}c}
      \hline \hline
      Symbol & {$\Tp$} & $\ro$ & $\Rey$ & $\Pep$ & $\rsup$ & $\rrms$ & $\urms$ & $b$\\ \hline 
       --&--     &--    & 100 & --  &--          & --                           & $2.42 \pm	0.05$ & --\\ 
       --&--     &--    & 300 & --  &--         & --                           & $2.43 \pm	0.16$ & -- \\
       --&--     &--    & 600 & --  &--        & --                           & $2.50 \pm 0.18$ & --\\ \hline
         \includegraphics[scale=0.5]{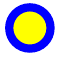}         &0.005 	&	0.1	& 100	& 1000	& $2.04 \pm 0.11$ 		&	$0.23	 \pm 0.005$ &	$2.17 \pm 0.08$ & $10.92 \pm 0.18$ \\
        \includegraphics[scale=0.5]{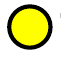}           &0.005 	&	0.1	& 100	& 600		& $1.92 \pm 0.11$  	& $0.22 \pm	0.01$ 	&	$2.28 \pm	0.07$ & $18.04 \pm 0.05$\\
        \includegraphics[scale=0.5]{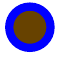}          &0.005 	& 	0.1 	&	600	&	1000	&	$3.84 \pm 0.43$ 	&	$0.29 \pm	0.01$	&	$2.26 \pm	0.08$ & $8.14 \pm 0.01$ \\
        \includegraphics[scale=0.5]{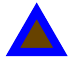}            &0.005	&	1		& 600	&	1000	&	$2.35 \pm	0.18$	& $0.21	\pm 0.01$	&	$1.80 \pm	0.09$ & $14.79 \pm 0.03$ \\
        \includegraphics[scale=0.5]{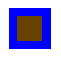}           &0.005	&	10		&	600	&	1000	&	$1.27 \pm	0.03$	&	$0.06 \pm	0.004$ 	&	$0.71 \pm	0.01$ & $51.72 \pm 0.85$ \\
      \hline
      \includegraphics[scale=0.5]{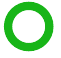}&0.01			&0.1	&	100	&	100		&	$1.85 \pm	0.18$	&	$0.22 \pm	0.03$	&	$2.20 \pm	0.06$ & $13.65 \pm 0.05$\\
      \includegraphics[scale=0.5]{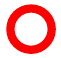}&0.01			&0.1		&100	&	300		&	$2.56 \pm	0.35$	&	$0.31 \pm	0.03$	&	$2.20 \pm	0.09$ & $9.25 \pm 0.01$\\
      \includegraphics[scale=0.5]{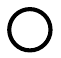}&0.01			&0.1	&	100	&	600		&	$3.17 \pm	0.36$	&	$0.37 \pm	0.03$	&	$2.22 \pm	0.025$ &$10.18 \pm 0.01$\\
      \includegraphics[scale=0.5]{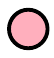}&0.01			&0.1	&	300	&	600		&	$5.61 \pm	1.21$	& $0.41 \pm	0.014$	&	$2.31 \pm	0.07$ & $5.58 \pm 0.004$\\
      \includegraphics[scale=0.5]{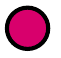}&0.01			&0.1		&600	&	600		&	$6.36 \pm	1.14$	&	$	0.40 \pm	0.01$	&	$2.16 \pm	0.056$ & $5.52 \pm 0.007$\\
      \includegraphics[scale=0.5]{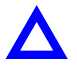}&0.01			&	1		&	100	&	1000	&	$	2.40\pm	0.27$	&	$0.23 \pm	0.02$	&	$1.58 \pm 	0.025$ & $11.60 \pm 0.025$ \\
      \includegraphics[scale=0.5]{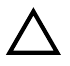}&0.01			&	1		&	100	&	600		& 	$2.10\pm	0.17$	& $0.23	\pm 0.01$	&	$1.63 \pm	0.02$ & $13.96 \pm 0.03$\\
      \includegraphics[scale=0.5]{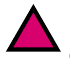}&0.01			&1		&600	& 600		&	$2.93 \pm	0.22$	&	$	0.26 \pm	0.01$	&	$	1.69 \pm	0.02$ & $9.32 \pm 0.01$ \\
      \includegraphics[scale=0.5]{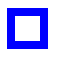}&0.01			&	10		&100	&1000		&  $1.53 \pm	0.08$	&	$0.10 \pm	0.006$		&	$0.82 \pm	0.05$ & $33.91 \pm 0.21$ \\
      \includegraphics[scale=0.5]{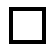}&0.01			&10		&100	&600		&	$1.31	\pm 0.03$ &	$0.09 \pm 0.002$ &	$0.69	\pm 0.004$ & $31.77 \pm 0.30$\\
      \includegraphics[scale=0.5]{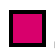}&0.01			&10		&600	&600		&	$1.39 \pm	0.05$	&	$0.08 \pm	0.002$ &	$0.69 \pm	0.002$ & $45.47 \pm 0.48$ \\
      \hline
      \includegraphics[scale=0.5]{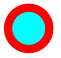}&0.03		    &0.1		&100	&300		&	$9.81\pm	2.50$ 		&	$0.64 \pm 0.03$	& $2.16 \pm	0.06$& $2.50 \pm 0.003$ \\
	  \includegraphics[scale=0.5]{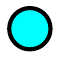}&0.03		    &0.1		&100	&	600		&	$8.46\pm	1.81$	&	$0.61 \pm	 0.014$	&	$2.05 \pm	0.07$ & $2.92 \pm 0.004$\\
	  \includegraphics[scale=0.5]{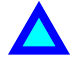}&0.03		    &1		&100	&1000		& $6.95 \pm 0.81$ 	& $0.49 \pm	0.02$	& $1.57	\pm 0.03$& $3.36 \pm 0.01$ \\
             \hline
      \end{tabular}
  \caption{Characteristics of the numerical simulations. The first column represents the markers for  Figures \ref{fig:max_enh}, \ref{fig:typ_enh}, and \ref{fig:bslope}. The second to fifth columns show $\Tp, \ro, \Rey,$ and $\Pep$ (where we have set $\Rep = \Pep$). The sixth to eighth columns show temporally averaged values for $\rsup$, $\rrms$, and $\urms$ once the system has reached a statistically steady state, and the errors represent a standard deviation around the mean. The last column corresponds to the slope $b$ and its standard error of the exponential tail of the pdfs presented in Section \ref{subsec:typ_enh}. All 3D simulations were run with $(L_x \times L_y \times L_z) = (4\pi \times 2\pi \times 2\pi)$ with the corresponding number of mesh points used in each direction as $(N_x \times N_y \times N_z) = (768 \times 384 \times 384)$. }
  \label{table:sims}
  \end{center}
\end{table}

\subsection{Quantifying particle concentration enhancement}\label{subsec:quant_partenhancement}

In what follows, we quantify particle concentration enhancement by first defining several terms: 
\begin{equation}\label{eqn:rsup}
    \rsup(t) = \max_{\mathbf{x}} \rhat(\mathbf{x},t) 
\end{equation}
representing the maximum particle concentration across the domain,  
\begin{equation}\label{eqn:rrms}
    \rrms(t) = \bigg[ \frac{1}{L_x L_y L_z} \iiint (\rhat - 1)^2 \hspace{2pt} dx dy dz\bigg] ^{1/2},
\end{equation}
defined as the standard deviation around the mean particle density $\rhat = 1$, and the rms fluid velocity defined as
\begin{equation}\label{eqn:urms}
    \urms(t) = \bigg[ \frac{1}{L_x L_y L_z} \iiint [{\uhat^2(\mathbf{x}, t)} + {\vhat^2(\mathbf{x}, t)} +  {\what^2(\mathbf{x}, t)}] \hspace{2pt} dx dy dz \bigg]^{1/2}.
\end{equation}

\begin{figure}
    \includegraphics[width=1\textwidth]{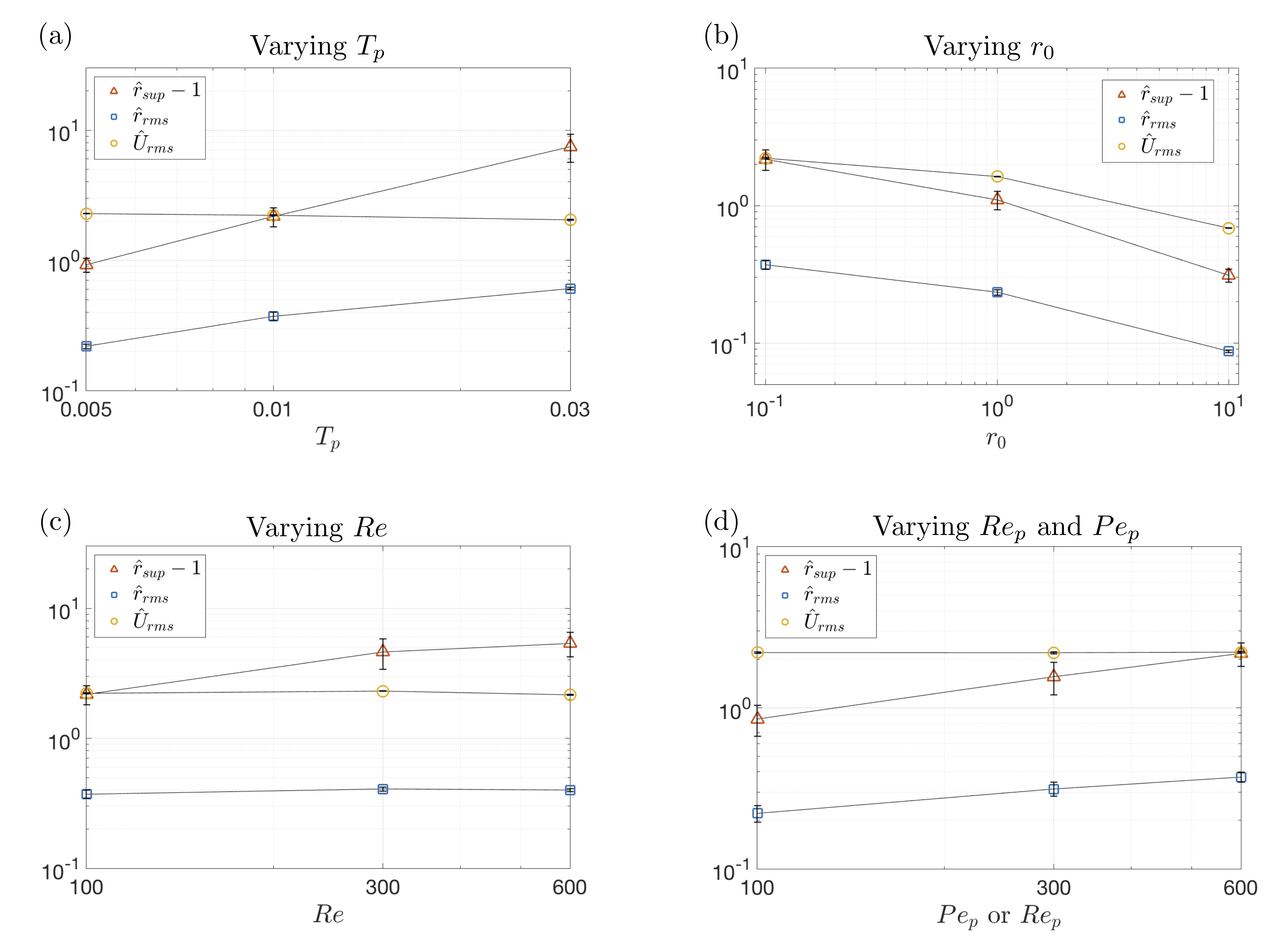}
        \caption{Temporally averaged maximum and typical particle concentration enhancement ($\rsup - 1$ and $\rrms$, respectively) and the temporally averaged rms fluid velocity $\urms$ for (a) varying $\Tp$, (b) varying $r_0$, (c) varying $\Rey$, and (d) varying $\Pep$ and $\Rep$ from simulations that have reached a statistically steady state. Error bars represent one standard deviation around the mean. Unless otherwise stated, $\Tp = 0.01, \ro =0.1$, $\Rey =100$, $\Pep = 600$, and $\Rep = 600$ (see main text). More details of the simulations can be found in Table \ref{table:sims}. }
        \label{fig:comparetpr0}
\end{figure}

We first look at how the two measures of particle concentration enhancement defined above, as well as the rms fluid velocity, vary with respect to $\Tp$, $\ro$, $\Rey$, and $\Pep$ (assuming as above that $\Rep=\Pep$). For each simulation presented, we take a temporal average of the quantities defined by Eqs. \eqref{eqn:rsup} -- \eqref{eqn:urms} after the system has reached a statistically steady state over a time range $\Delta t$, and report the means as $\rsup, \rrms,$ and $\urms$ in Table \ref{table:sims}. We then take the standard deviation around this temporal average as an estimate of the errorbar (quantifying the variability). Figure \ref{fig:comparetpr0} presents the temporally averaged values of $\rsup - 1$, $\rrms$, and $\urms$ for selected simulations. In  Figure \ref{fig:comparetpr0}(a), we present simulations for varying $\Tp$, while holding $\ro = 0.1, \Rey = 100, \Rep = 600,$ and $\Pep = 600$ constant. We see that both $\rsup -1$ and $\rrms$ increase with $\Tp$, while $\urms$ is overall unaffected. This is consistent with the observation in Section \ref{subsec:compare_tp} that $\Tp$ only has a small effect on the overall power spectrum of the turbulence, but directly controls the rate of preferential concentration. In  Figure \ref{fig:comparetpr0}(b), $\ro$ is varied, while $\Tp = 0.01,  \Rey = 100, \Rep = 600,$ and $\Pep = 600$ are held constant. We see that all quantities decrease with increasing $\ro$. This can be explained by the fact that an increase in $\ro$ corresponds to a decrease in the turbulent fluid velocity, resulting in a subsequent decrease in the particle concentration enhancement (see Section \ref{subsec:compare_ro}). Moving on to  Figure \ref{fig:comparetpr0}(c) where $\Rey$ is varied while $\Tp = 0.01, \ro = 0.1, \Rep = 600,$ and $\Pep = 600$ are held constant, we see that $\urms$ and $\rrms$ are overall unchanged, at least within the range of $\Rey$ shown. In contrast, we see a slight increase of $\rsup - 1$ with $\Rey$. Finally in  \ref{fig:comparetpr0}(d), where $\Pep$ (and $\Rep$) is varied while $\Tp = 0.01, \ro = 0.1,$ and $\Rey = 100$ are held constant, we see that both $\rsup -1$ and $\rrms$ increase with $\Pep$, while $\urms$ is unchanged (see Section \ref{subsec:compare_Pep_Rep}). Therefore, we see that the quantities $\rsup$, $\rrms$, and $\urms$ for varying parameters are consistent with the spectra shown in Sections \ref{subsec:compare_tp}-\ref{subsec:compare_Pep_Rep}.

\section{Predictive model}\label{sec:predictive_model}

As discussed in Section \ref{sec:intro}, \citet{nasab2020preferential} found that the maximum particle concentration in a fluid where the turbulence is driven by the particle Rayleigh-Taylor instability scales as $u^2_{rms} \tau_p/ \kappa_p$, and presented theoretical arguments of dominant balance that support this law. For pedagogical purposes, we reproduce the arguments here, and then verify whether the same scaling law applies for particles in mechanically-driven (shear-induced) turbulence as studied in this paper. 

We start with the particle concentration equation \eqref{eqn:nd_part} and substitute $\rhat = 1 + \rhat'$ to get
\begin{equation}
    \frac{\partial \rhat'}{\partial t} + (1 + \rhat')\div \bup + \bup \cdot \grad \rhat' = \frac{1}{\Pep} \nabla^2 \rhat',
\end{equation}
where $\rhat'$ is the particle concentration enhancement over the mean. 

As in \citet{nasab2020preferential}, we assume that in regions of maximal concentration enhancement there is a dominant balance between the inertial concentration term and the diffusion term expressed as
\begin{equation}\label{eqn:dombal}
    \div \bup \sim \frac{1}{\Pep} \nabla^2 \rhat'.
\end{equation}
Using Eq. \eqref{eqn:divup} in Eq. \eqref{eqn:dombal}, we obtain
\begin{equation}
    -\Tp \div (\bu \cdot \grad \bu) \sim \frac{1}{\Pep}\nabla^2 \rhat'.
\end{equation}

Assuming that the characteristic lengthscale involved in the inertial term and the diffusive term are the same, dimensional analysis reveals that 
\begin{equation}\label{eqn:scalingmax}
    \rhat' \sim \urms^2 \Tp \Pep,
\end{equation}
where $\urms$ represent the characteristic fluid velocity of the system (see Eq. \ref{eqn:urms}). Dimensionally, this expression becomes 
\begin{equation}\label{eqn:scalingmax_dim}
    \bigg(\frac{\rho_p'}{\bar{\rho}_p}\bigg)_{max} \sim \frac{u^2_{rms} \tau_p}{\kappa_p},
\end{equation}
where $\rho'_p = \rhat' r_0 \rho_s$ is the local particle density enhancement over the mean $\bar{\rho}_p = r_0 \rho_s $ (see \cite{nasab2020preferential} for more details) and $u_{rms}$ is the dimensional rms fluid velocity. We now have a scaling law relating particle concentration enhancement $\rhat'$ to only three properties of the flow: the characteristic fluid velocity, the particle stopping time, and the particle diffusivity. 

\subsection{Maximum particle concentration enhancement}\label{subsec:max_enh}

\begin{figure}
    \centering
        \includegraphics[width=0.75\textwidth]{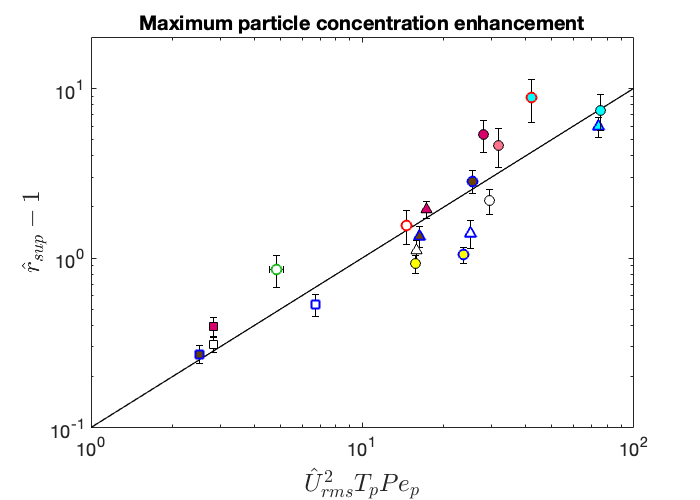}
        \caption{Maximum particle concentration enhancement over the mean as function of $\urms^2 \Tp \Pep$ with varying parameters (i.e. $\Tp$, $\ro$, $\Rey$, $\Rep$, and $\Pep$). The black solid line represents $ \rsup -1 = (1/10)\urms^2 \Tp \Pep$. The legend and details of simulations can be found in Table \ref{table:sims}.  }
        \label{fig:max_enh}
\end{figure}

As in \citet{nasab2020preferential}, we compare the scaling law (\ref{eqn:scalingmax}) to our selected measure of maximum particle concentration enhancement $\rsup - 1$ (see Eq. \ref{eqn:rsup}). In  Figure \ref{fig:max_enh}, we present $\rsup - 1$ versus $\urms^2 \Tp \Pep$ (with the legend and simulation details found in Table \ref{table:sims}). Each point corresponds to one simulation, where the values of $\rsup$ and $\urms$ were extracted after the system has reached a statistically steady state. Various marker types represent varying $\Tp$, $\ro$, $\Rey$, $\Pep$, and $\Rep$: the color represents the value for $\Tp$ or $\Rey$, the shape represents $\ro$, and colored outlines represent $\Pep$ (or equivalently $\Rep$, since $\Rep = \Pep$). The solid line represents the scaling $\rhat' \sim \urms^2 \Tp \Pep$. 

Our main conclusion is that the scaling law proposed by \citet{nasab2020preferential} in the context of the particle-driven convective instability 
also holds more generally in mechanically-driven turbulent flows, which is perhaps not surprising, but needed to be established. As expected, we see points for larger $\Tp$ or smaller $\ro$ have larger $\rsup$, while larger $\ro$ results in a smaller $\urms$, and therefore smaller $\rsup$, as discussed in Section \ref{subsec:quant_partenhancement}. We also see that for larger $\Rey$, $\urms$ increases slightly, resulting in larger $\rsup$.

\subsection{Typical particle concentration enhancement}\label{subsec:typ_enh}

\begin{figure}
    \centering
        \includegraphics[width=0.75\textwidth]{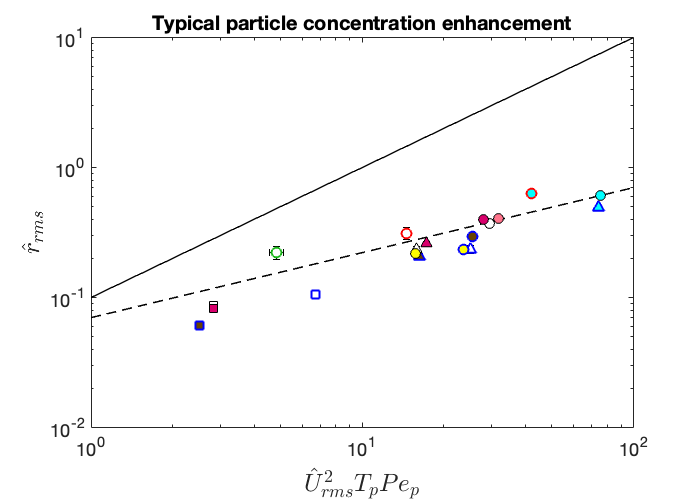} 
        \caption{Typical particle concentration enhancement over the mean as a function of $\urms^2 \Tp \Pep$ with varying parameters (i.e. $\Tp$, $\ro$, $\Rey$, $\Rep$, and $\Pep$). The black dotted line represents $\rrms = (0.07) \urms (\Tp \Pep)^{1/2}$ and the solid line represents $\rrms  = (1/10)\urms^2 \Tp \Pep$. The legend and details of simulations can be found in Table \ref{table:sims}.}
        \label{fig:typ_enh}
\end{figure}

In our previous work \cite{nasab2020preferential}, we also showed that the typical particle concentration enhancement $\rrms$ did not follow the scaling law given by Eq. \eqref{eqn:scalingmax}, but instead scaled like 
\begin{equation}
    \rrms \sim \sqrt{\urms^2 \Tp \Pep},
\end{equation}
which dimensionally is 
\begin{equation}\label{eqn:scalingtyp_dim}
    \bigg( \frac{\rho_p'}{\bar{\rho}_p} \bigg)_{rms} \sim  \bigg( \frac{u^2_{rms} \tau_p}{\kappa_p} \bigg)^{1/2}.
\end{equation}
We see that this result also holds for this work in  Figure \ref{fig:typ_enh}. The data points do not follow the scaling law \eqref{eqn:scalingmax} shown by the solid line, and instead follow the dashed line representing $\rrms \sim \urms \sqrt{\Tp \Pep}$.

\begin{figure}
    \includegraphics[width=1\textwidth]{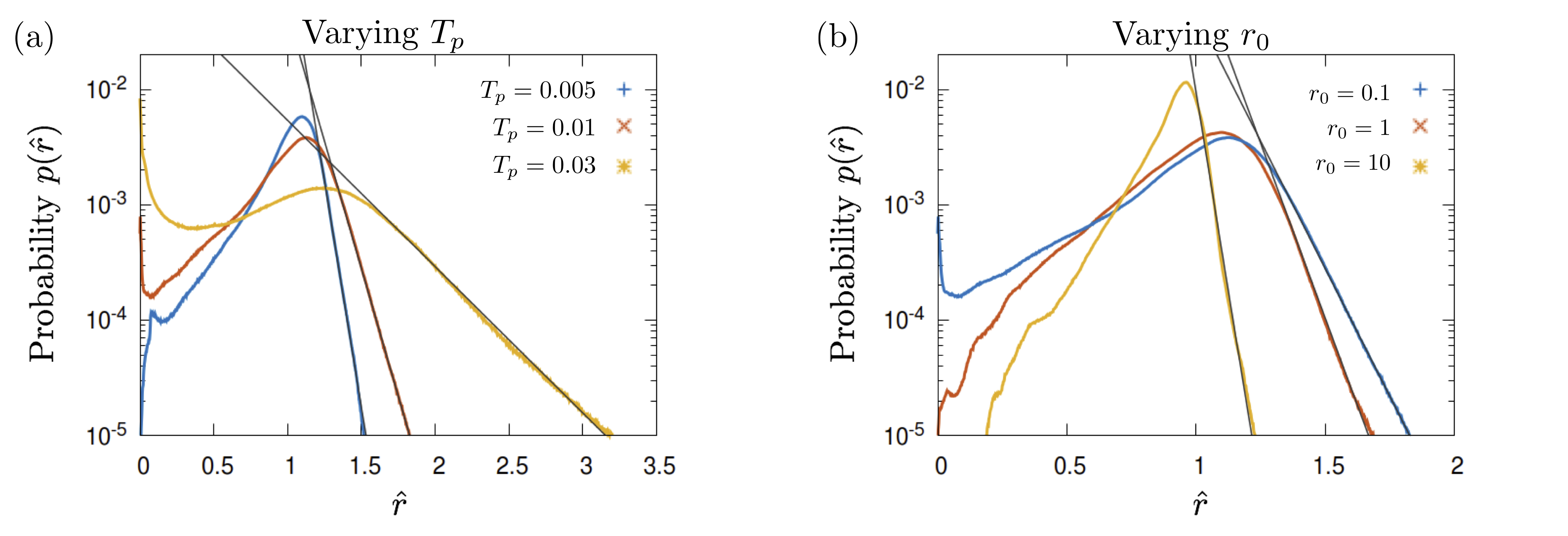}    
        \caption{Probability distribution functions for $\rhat$, computed from simulations that have reached a statistically steady state (a) for varying $\Tp$ with $\ro = 0.1, \Rey = 100, \Pep = 600, \Rep=600$ and (b) for varying $\ro$ with $\Tp = 0.01, \Rey = 100, \Pep = 600, \Rep =600$. The gray lines fit the tail of each pdf and are of the form $p(\rhat) \propto e^{-b\rhat}$. 
        Values of $b$ and simulation details can be found in Table \ref{table:sims}.}
        \label{fig:pdf}
\end{figure}

As argued by \citet{nasab2020preferential}, the fact that $\rrms$ depends on the same {\it combination} of parameters as $\rsup-1$ (albeit with a different power law) strongly suggests that the entire pdf of the concentration enhancement depends on the combination $\urms^2 \Tp \Pep$. To see whether a similar argument applies here, Figure \ref{fig:pdf} presents pdfs of $\rhat$ for selected simulations of varying (a) $\Tp$ and (b) $\ro$ (with simulation details in Table \ref{table:sims}). These pdfs represent the probability of one pixel in the simulation to have a concentration whose value lies between $\rhat$ and $\rhat + \Delta \rhat$, where $\Delta \rhat = 0.002$. The pdf would take the form of a delta function centered on $\rhat = 1$ in the absence of preferential concentration ($\Tp \rightarrow 0$), since the particle density in that case remains equal to one everywhere and at all times.


 On the other hand when preferential concentration is present, the pdf broadens as the particle density becomes more inhomogeneous. We see in  Figure \ref{fig:pdf}(a), where $\Tp$ is varied while holding $\ro = 0.1, \Rey = 100, \Pep =600$, and $\Rep = 600$ constant, that 
 the pdf appears relatively narrow around the mean value $\rhat = 1$ for small $\Tp$. As $\Tp$ increases, the spatial distribution of the particles becomes more heterogeneous due to preferential concentration, and the pdf widens considerably. We also see an increase in the probability of events of no particles (when $\rhat \simeq 0$) and the appearance of an elongated tail capturing extreme events where the particle concentration is largest. The tail is exponential, of the form $p(\rhat)\propto e^{-b \rhat}$ (see \citet{nasab2020preferential}). 
 
 Moving to  Figure \ref{fig:pdf}(b) in which $\ro$ is varied while $\Tp = 0.01, \Rey =100, \Pep =600$, and $\Rep = 600$, we see that increasing $\ro$ causes the pdf to become narrower. This is consistent with the fact that a larger $\ro$ lowers the amplitude of the turbulence in the system, and consequently, weakens preferential concentration. 

\begin{figure}
    \centering
        \includegraphics[width=.6\textwidth]{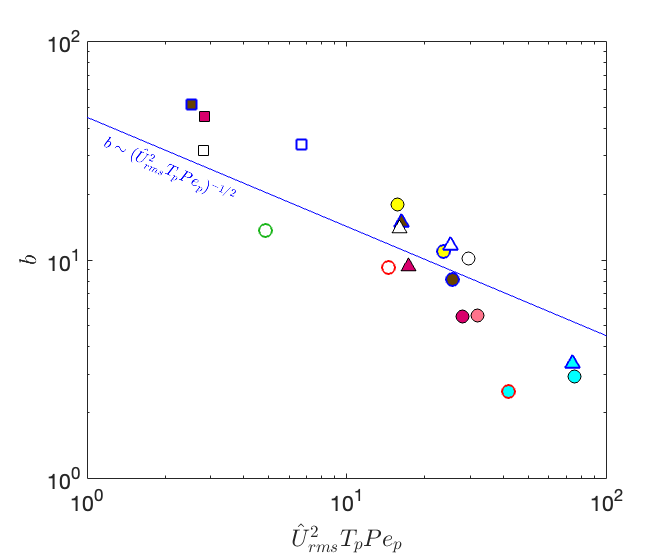} 
        \caption{The slope $b$ of the exponential tail of the pdf of  $\rhat$ as a function of $\urms^2 \Tp \Pep$ for simulations at various $\Tp$, $\ro$, $\Rey$, $\Rep$, and $\Pep$ (where $\Rep = \Pep$). The blue solid line shows $b \sim (\urms^2 \Tp \Pep)^{-1/2}$. See Table \ref{table:sims} for more details. }
        \label{fig:bslope}
\end{figure}

\citet{nasab2020preferential} studied more quantitatively the exponential tail of the pdf, which seems to be almost ubiquitous, and found that its decay rate $b$ scales as $(\urms^2 \Tp \Pep)^{-1/2}$. In Figure \ref{fig:bslope}, we  present $b$ as a function of $\urms^2 \Tp \Pep$, where $b$ was found by fitting a decaying exponential function to the pdfs 
presented in  Figure \ref{fig:pdf}, along with additional pdfs computed from simulations with varying $\Rey$, $\Pep$, and $\Rep$. Each simulation is represented by one data point with the same marker type used
in  Figures \ref{fig:max_enh} and \ref{fig:typ_enh} (with simulation details in Table \ref{table:sims}), where the errors on $b$ are not shown since they are much smaller than the marker size. The data points appear to follow the blue line given by  $(\urms^2 \Tp \Pep)^{-1/2}$, consistent with results from \citet{nasab2020preferential}. This demonstrates that $\rrms \sim 1/b$, a result that is not entirely surprising since it would actually be exact if the pdfs were purely exponential.

\section{Summary, Applications, and Discussion}\label{sec:summary}
\subsection{Summary} \label{subsec:summary}

In \citet{nasab2020preferential} we studied preferential concentration in a two-way coupled particle-laden flow in the context of the particle-driven convective instability. We found that the maximum particle concentration enhancement above the mean scales as $u_{rms}^2 \tau_p / \kappa_p$, where $u_{rms}$ is the rms fluid velocity, $\tau_p$ is the particle stopping time, and $\kappa_p$ is the assumed particle diffusivity. Additionally, we found that the typical particle concentration enhancement over the mean scales as $(u_{rms}^2 \tau_p / \kappa_p)^{1/2}$ and the pdf of the particle concentration over the mean has an exponential tail whose slope scales like $(u_{rms}^2 \tau_p /\kappa_p)^{-1/2}$. However, it was not clear that these results would remain
valid in a system in which the turbulence is not driven by the particles themselves. In this work, we confirm that the results of \citet{nasab2020preferential} apply in 
a system in which the turbulence is mechanically-driven. With this extension to a much wider class of turbulent systems, our model has important consequences for preferential concentration in the various applications introduced in Section \ref{sec:intro}. In the next sections, we discuss the potential caveats one should bear in mind before applying the model to real physical systems, and  present a particular prediction of the model for droplet concentration in clouds.

\subsection{Extension of the model to higher $\Rey$} \label{subsec:discussion}


In general, realistic applications of preferential concentration in natural systems take place in environments such as clouds, river outflows, or accretion disks, that are highly turbulent in nature, and whose Reynolds numbers are asymptotically large. Because of this, the  
velocity spectra have an inertial range which spans many orders of magnitudes in lengthscales. For sufficiently turbulent flows, it is well established that the Stokes number increases with wavenumber 
and reaches a maximum at the end of the inertial range (i.e. near the Taylor microscale). This poses two problems in terms of the extension of our results to very strongly turbulent flows. On the one hand, it is possible for the Stokes number at the Taylor microscale to exceed the threshold of validity of the two-fluid approximation (even if $St < 0.3$ at the injection scale). In that case, our results are not expected to apply. On the other hand, even if the two-fluid approximation remains valid, one could question whether our results, which were obtained for moderate Reynolds numbers, still apply when $\Rey \rightarrow \infty$.  

Indeed, due to the high computational cost of running 3D DNSs, we only looked at moderately turbulent systems where $\Rey \lesssim 600$. In that case,
the inertial range of the velocity spectra is 
quite limited (see Section \ref{sec:num_sims}). As a result, the characteristic fluid velocity measured at the injection scale is
comparable to the corresponding fluid velocity measured at the Taylor microscale. However when $\Rey \rightarrow \infty$, the velocities at these two scales may be vastly different. This naturally brings up a valid question concerning the predictive model: is the maximum particle concentration enhancement dependent on the fluid velocity measured at the injection scale (as we assumed in this work), or at the Taylor microscale where the Stokes number is maximal? A further look into the data may provide some preliminary clue to the answer (although simulations at much higher $\Rey$ will ultimately be needed to fully confirm the results).

We saw in Section \ref{sec:predictive_model} that 
our predictive model for maximum particle concentration enhancement in the two-fluid approximation depends
on the fluid velocity, the particle stopping time, and the assumed particle diffusivity, as
\begin{equation}\label{eqn:model_disc_general}
	\bigg( \frac{\rho'_p}{\bar{\rho}_p} \bigg)_{max} \approx \alpha \frac{u^2(\ell) \tau_p}{\kappa_p}
\end{equation}
for some prefactor $\alpha$, where here we allow for the possibility that the correct value of $u$ may be different from $u_{rms}$. 
We now consider the hypothesis raised above that 
the fluid velocity may need to be that of the Taylor microscale instead, such that $\ell = \lambda$. Assuming a Kolmogorov spectrum for the kinetic energy, it then follows that 
\begin{equation}\label{eqn:u_ell}
	u(\lambda) = u_{rms} (\lambda k_s )^{1/3}, 
\end{equation}
where we recall that $2\pi/k_s=L_z$ is the height of the computational domain. 
Using the fact that 
$\lambda k_s = \sqrt{15}\Rey^{-1/2}$ 
we obtain 
\begin{equation}\label{eqn:u_taylormic}
	u(\lambda) = u_{rms} (15)^{1/6} \Rey^{-1/6}.
\end{equation}
Substituting $u(\lambda)$ in \eqref{eqn:model_disc_general}, the maximum particle concentration enhancement in this alternative model would be
\begin{equation}
	\bigg( \frac{\rho_p'}{\bar{\rho}_p} \bigg)_{max} \sim \Rey^{-1/3} \frac{u^2_{rms} \tau_p}{\kappa_p}.
\end{equation}
 
This formula suggests that $(\rho_p'/\bar{\rho}_p)_{max}$ should decrease with increasing $\Rey$. If this were the case, then we would expect that $(\rho_p'/\bar{\rho}_p)_{max}$ should be approximately twice as large
for $\Rey = 100$ in comparison to $\Rey = 600$ (with the remaining parameters fixed to be the same). This is contrary to the observations from our simulations, in which we see 
the opposite trend (see, e.g. Figure \ref{fig:comparetpr0}(c)). 
We therefore conclude that our original model, in which $(\rho_p'/\bar{\rho}_p)_{max} \sim u_{rms}^2 \tau_p/\kappa_p$, is more likely to be correct. These results, however, will need to be confirmed more directly with simulations at much higher Reynolds number.

\subsection{Application to natural systems}\label{subsec:implications}

While the question of the applicability of our model to very large Reynolds number systems was partially addressed in the previous section, a second, much more difficult question arises concerning the applicability and validity of the two-fluid equations themselves. In particular, the central result of this work is the role of particle diffusion ($\kappa_p$) in controlling the maximum and typical (rms) particle concentration enhancement (see Sec. \ref{subsec:max_enh}-\ref{subsec:typ_enh}), so one may rightfully question whether the diffusion approximation used in Eqs. \eqref{eqn:mom}--\eqref{eqn:divfree} is valid in the first place. A complete answer to this question is largely beyond the scope of this paper, and will require either delicate laboratory experiments, or DNSs of a large number of fully-resolved particles interacting with a turbulent fluid. 

In the limit where the particles are very small, however, stochastic collisions with the fluid molecules are a source of dispersion in the particle transport equation (usually referred to as Brownian motion), that can be modeled as a diffusion process and whose coefficient is given by 
\begin{equation}
    \kappa_p \approx \frac{k_B T_m}{6 \pi s_p \rho_f \nu},
\end{equation}
where $k_B = 1.38 \times 10^{-23}$ J $\cdot$ K$^{-1}$ is the Boltzmann constant, $T_m$ is the mean temperature of the fluid, and $s_p$ is the particle radius. This expression can be considered as a lower limit on the effective particle diffusivity, and using it in conjunction with Eq. \eqref{eqn:scalingmax_dim} and  \eqref{eqn:scalingtyp_dim}, provides an upper limit on the maximum particle concentration  $(\rho_p'/\bar{\rho}_p)_{max}$ and the rms particle concentration enhancement $(\rho_p'/\bar{\rho}_p)_{rms}$.

To see what kind of prediction for particle concentration this lower-limit estimate for $\kappa_p$ leads to, it is helpful to consider a specific application, such as that of rain formation in warm clouds (e.g. cumulus or stratocumulus clouds). In this application, turbulence is generally mechanically-driven, generated by vertical drafts and wind shear. It has been largely hypothesized that the broadening of the droplet spectrum during the initial stage of droplet growth is due to preferential concentration followed by enhanced collision rates and coalescence \cite{shaw2003particle}. With this in mind, we consider small droplets of radius $s_p = 10$ $\mu$m and density $\rho_s = 1000$ kg/m$^3$ with the typical values for the properties of ambient air being $\rho_f = 1$ kg/m$^3$, $\nu \approx 10^{-5}$ m$^2$/s, and a mean temperature of $T_m \approx 300$ K.

Based on these estimates, the stopping time for a cloud droplet is given by 
\begin{equation}
	\tau_p = \frac{2 \rho_s s_p^2}{9 \rho_f \nu} \approx (2 \times 10^{-3} {\text { s}}) \bigg( \frac{s_p}{10 \text{ }\mu \text{m}} \bigg) ^2 ,
\end{equation}
and the particle diffusivity due to Brownian motion is given by
\begin{equation}
    \kappa_p = (2 \times 10^{-12} \text{ m}^2\text{/s}) \bigg( \frac{10 \text{ } \mu\text{m}}{s_p} \bigg) \bigg( \frac{T_m}{300 \text{ K}} \bigg). 
\end{equation}

Using this, we can then obtain an upper limit estimate of the maximum and rms particle concentration enhancements as 
\begin{align}
	& \bigg( \frac{\rho_p'}{\bar{\rho}_p} \bigg)_{max} \lesssim \alpha \frac{u_{rms}^2 \tau_p}{\kappa_p} \approx 10^8 \bigg(\frac{u_{rms}}{\text{1 m/s}} \bigg)^2 \bigg(\frac{s_p}{10 \text{ }\mu \text{m}} \bigg)^2 \bigg( \frac{2 \times 10^{-12} \text{ m}^2\text{/s}}{\kappa_p} \bigg) \label{eqn:maxenh_brownian} \\
	& \bigg( \frac{\rho_p'}{\bar{\rho}_p} \bigg)_{rms} \lesssim \gamma u_{rms} \sqrt{\frac{\tau_p}{\kappa_p}} \approx 2 \times 10^3 \bigg(\frac{u_{rms}}{\text{1 m/s}} \bigg) \bigg(\frac{s_p}{10 \text{ }\mu \text{m}} \bigg) \bigg(\frac{2 \times 10^{-12} \text{ m}^2/\text{s}}{\kappa_p} \bigg)^{1/2}
\end{align}
where we have used $\alpha \approx 0.1$ and $\gamma \approx 0.07$ extracted from our simulations (see Figures \ref{fig:max_enh} and \ref{fig:typ_enh}), and a fiducial value of $u_{rms} = 1$ m/s was assumed. This result is quite remarkable, given that the characteristic Stokes number $\St$ associated with these droplets is very small. Indeed, assuming that the eddy turnover time is $\tau_e \sim L/u_{rms}$ where $L \sim 1$ km is a typical cloud height, we find that  
\begin{equation}
    St \simeq \frac{\tau_p}{\tau_e} \approx (2 \times 10^{-6})\bigg(\frac{s_p}{10 \text{ }\mu \text{m}} \bigg)^2 \bigg( \frac{1 \text{ km}}{L} \bigg) \bigg(\frac{u_{rms}}{\text{1 m/s}} \bigg).
\end{equation}
This suggests that strong preferential concentration is possible even when $St \ll 1$ (a surprising result that is supported by the DNSs presented in Section \ref{sec:num_sims}).

Of course, as discussed above, this provides only an upper limit estimate of the particle concentration enhancement, which is only valid as long as $\kappa_p$ is dominated by the effects of Brownian motion. To check whether this is likely true in the cloud application considered, we compute the  corresponding volume fraction occupied by the particles in regions of maximal concentration. We find that if the mean liquid water content of the cloud is $\bar \rho_p \approx 1$ mg/m$^3$, then the average volume fraction occupied by the droplets is $\bar{\Phi} = (\bar{\rho}_p/\rho_s) \approx 10^{-9}$. Thus, the associated maximum and typical volume fraction achievable though preferential concentration are 
\begin{align}
&\Phi_{max} \approx \bar{\Phi} ( 
{\rho_p'}/{\bar{\rho}_p} )_{max} \approx O(0.1), \\
&\Phi_{rms} \approx \bar{\Phi} ( {\rho_p'}/{\bar{\rho}_p} )_{rms} 
\approx O(10^{-6}).
\end{align}

With the possibility of very large volume fractions emerging out of the preferential concentration process, we must therefore account for the possibility that particles may interact hydrodynamically through their wakes, which would increase $\kappa_p$ (and therefore lower $( {\rho_p'}/{\bar{\rho}_p} )_{max}$ and $\Phi_{max}$, and possibly also $( {\rho_p'}/{\bar{\rho}_p} )_{rms}$ and $\Phi_{rms}$). For simplicity, we use the results of  \citet{segre2001effective} to construct an effective diffusion coefficient associated with hydrodynamic interactions. They suggest that that the mutually-induced dispersion can be modeled by 
\begin{equation}
    \kappa_p \approx \beta(\Phi) s_p V_p,
    \label{eq:kappahydro}
\end{equation}
where $\beta$ is a function of the volume fraction $\Phi$ occupied by the particles and $V_p$ is the velocity of the particles relative to the fluid. \citet{segre2001effective} found that $\beta(\Phi) \lesssim 0.1$ for volume fractions of up to $\Phi \approx 0.2$. Thus we can construct 
an approximate upper limit for $\kappa_p$ by setting $\beta = 0.1$. 

The relative velocity of the particles with respect to the fluid 
is obtained following \citet{maxey_1987} (and the arguments presented in Eq. \ref{eqn:up_asymp}) to be 
\begin{equation}
	V_p = |\boldup - \boldu|  \approx \tau_p \bigg|  \frac{\partial \boldu}{\partial t} + \boldu \cdot \grad \boldu \bigg| + O(\tau_p^2 ) .
\end{equation}
We can estimate it roughly using dimensional arguments as
\begin{equation}
	V_p(\ell) \approx \tau_p \frac{u^2(\ell)}{\ell} \approx \tau_p \frac{u^2_{rms}}{L} \bigg( \frac{\ell}{L} \bigg) ^{-1/3}, \label{eqn:vp_intermed}
\end{equation}
assuming a Kolmogorov scaling for the eddy velocity $u(\ell)$ at scale $\ell$. We therefore see that $V_p$ will be largest at the Taylor microscale, and set $\ell = \lambda \approx \sqrt{15} \Rey^{-1/2}L$ to obtain an upper limit for $V_p$:	 
\begin{equation}
	 V_p \lesssim \tau_p \frac{u^2_{rms}}{L} (15)^{-1/6}\Rey^{1/6} \label{eqn:vp_final}.
\end{equation}
Using \eqref{eqn:vp_final} in \eqref{eq:kappahydro} we can now obtain an upper limit on $\kappa_p$, as 
\begin{equation}
	\kappa_p \lesssim \beta s_p V_p \approx (3 \times 10^{-11} \text{ m}^2/\text{s})   \bigg(\frac{s_p}{10 \text{ }\mu \text{m}} \bigg)^3 \bigg(\frac{u_{rms}}{\text{1 m/s}} \bigg)^{13/6} \bigg( \frac{1 \text{ km}}{L} \bigg)^{5/6},
\end{equation}
which is only about one order of magnitude larger than the value for $\kappa_p$ obtained by considering the contribution due to Brownian motion only (for Eq. \ref{eqn:maxenh_brownian}).

We apply this formulation for $\kappa_p$ in \eqref{eqn:model_disc_general} and find that
	\begin{align}
	& \bigg( \frac{\rho_p'}{\bar{\rho}_p} \bigg)_{max} \lesssim \alpha \frac{u_{rms}^2 \tau_p}{\kappa_p} \approx 10^7 \bigg(\frac{u_{rms}}{\text{1 m/s}} \bigg)^2 \bigg(\frac{s_p}{10 \text{ }\mu \text{m}} \bigg)^2 \bigg(\frac{3 \times 10^{-11} \text{ m}^2/\text{s}}{\kappa_p} \bigg),  \\
	& \bigg( \frac{\rho_p'}{\bar{\rho}_p} \bigg)_{rms} \lesssim \gamma u_{rms} \sqrt{\frac{\tau_p}{\kappa_p}} \approx 600 \bigg(\frac{u_{rms}}{\text{1 m/s}} \bigg) \bigg(\frac{s_p}{10 \text{ }\mu \text{m}} \bigg) \bigg(\frac{3 \times 10^{-11} \text{ m}^2/\text{s}}{\kappa_p} \bigg)^{1/2},
\end{align}
with a corresponding maximal and rms volume fraction 
\begin{align}
    &\Phi_{max} \approx \bar{\Phi} ( {\rho_p'}/{\bar{\rho}_p} )_{max} 
    \approx O(0.01), \\
    &\Phi_{rms} \approx \bar{\Phi} ( {\rho_p'}/{\bar{\rho}_p} )_{rms} 
    \approx O(10^{-6}).
\end{align}
Note that since these were obtained using upper limits on $\kappa_p$, they can be viewed as lower limits on $\Phi_{max}$ and $\Phi_{rms}$.

Overall, this shows that both lower and upper limit estimates for the particle diffusivity $\kappa_p$ yield relatively consistent results in the context of cloud applications, and more importantly, that cloud turbulence could produce very large localized enhancements of the droplet concentration, despite the fact that the Stokes number is very low. Applications of this work to dust growth in protoplanetary disks were discussed by \citet{garaudnasab2019inertialaccretion}, with very similar conclusions. 
 
Of course, our results also show that these extreme events where $\Phi$ approaches $\Phi_{max}$ are rare, belonging to the tail of an exponential distribution. However, it is also well known in the context of both rain formation \cite{devenish2012droplet, grabowski2013growth} and planet formation \cite{birnstiel2016dust, weidenschilling1993formation}, that producing a few larger particles is all it takes for the process to start. Indeed, these larger ``lucky particles'' then sediment or drift with respect to the smaller ones, and can continue to grow by sweeping the latter. As such, particle growth in these contexts is controlled by what happens in the tail of the particle size distribution, which is why the results discussed here are particularly relevant.




\section*{Acknowledgements}
S. N. is supported by NSF AST-1908338. Simulations were run on a modified version of the PADDI code, originally written by S. Stellmach, on the UCSC Lux cluster and the NERSC Cori supercomputer. The authors thank Eckart Meiburg for his invaluable insight. 

\bibliographystyle{unsrtnat}
\bibliography{mybib.bib}

\end{document}